\documentstyle[epsfig]{jaa}
\begin{document}
\title[The Dynamics of Radio Galaxies]{The Dynamics of Radio Galaxies and Double-Double Radio Galaxies} 
\author[C. Konar]%
       {C. Konar$^1$\thanks{e-mail:chiranjib.konar@gmail.com}, M. Jamrozy$^2$, M.J. Hardcastle$^3$, 
J.H. Croston$^4$, \newauthor  S. Nandi$^5$, D.J. Saikia$^6$, J. Machalski$^2$  \\ 
$^1$ Institute of Astronomy and Astrophysics, AS, Taipei 10617, Taiwan. \\
$^2$ A.O.U.J., ul. Orla 171, 30244 Krak\'ow, Poland. \\
$^3$ University of Hertfordshire, College Lane, Hatfield, UK. \\
$^4$ University of Southampton, Southampton SO17 1BJ, UK. \\
$^5$ ARIES, Manora Peak, Nainital 263129, India. \\ 
$^6$ NCRA, TIFR, Pune University Campus, Pune 411007, India. \\   
}
\maketitle
\label{firstpage}
\begin{abstract}
Relativistic and magnetised plasma ejected by radio loud
AGNs through jets form the diffuse lobes of radio galaxies. The
radiating particles (electron/electron-positron) in lobes emit
in radio via the synchrotron process and X-ray via inverse-Compton
scattering of cosmic microwave background photons. The thermal
environment around radio galaxies emits X-rays via the thermal
bremsstrahlung process. By combining information from these processes
we can measure physical conditions in and around the radio lobes and
thus study the dynamics of radio galaxies, including double-double 
radio galaxies.
\end{abstract}

\section{Introduction}
\label{sec:intro}
Radio galaxies are created by Active Galactic Nuclei (AGNs) which eject jets in two opposite directions
and inflate what are called lobes. This jet forming activity of an AGN may be episodic in nature.
A striking example of episodic jet forming activity in radio galaxies is seen when a new pair of radio
lobes form closer to the nucleus before the `old' and more distant radio lobes have faded
(e.g. Subrahmanyan, Saripalli \& Hunstead 1996; Lara et al. 1999). Such radio galaxies have been
christened `Double-Double Radio Galaxies' (DDRGs) by Schoenmakers et al. (2000). Hereafter,
we will call the double lobed sources  `radio galaxies' and sources with two pairs of lobes with a
common centre (core) DDRGs (see Saikia \& Jamrozy (2009) for review).  

Lobes consist of charged particles which can be either electron-positron pairs or electron-proton pairs. 
The exact composition of lobes is not known.
Whatever may be the composition of lobes, it is well established that the lobes contain electrons 
(negatively charged particles) which radiate via the synchrotron and inverse-Compton (IC) processes. 
The synchrotron and IC power emitted by a single electron can be written as
\begin{equation}
P_{syn}=\frac{4}{3}c\sigma_T\gamma^2\beta^2U_{B}  
\end{equation}
and 
\begin{equation}
P_{IC}=\frac{4}{3}c\sigma_T\gamma^2\beta^2U_{ph}  
\end{equation}
(see Rybicki \& Lightman 1979)
where $c$ is velocity of light, $\beta$ is the velocity of the electron in units of $c$ and $\gamma$ is the 
corresponding Lorentz factor.
$U_B$ and $U_{ph}$ are the magnetic field energy density and photon field energy density respectively, in which the 
electron is moving. $\sigma_T$ is the Thomson cross section and is given by $\sigma_T= \frac{8\pi}{3}\frac{e^4}{c^4m_e^2}$, 
where $e$ is the charge of an electron and $m_e$ is the mass of an electron.  

Since there cannot be charge separation, there must be positively charged particles in lobes. If there are
positrons (as the positive component) then they will radiate (via synchrotron and IC processes) at the same 
rate as the electrons, as they have the same mass and magnitude of charge ($P_{syn} \propto e^4$ and
$P_{IC} \propto e^4$, where $P_{syn}$ is synchrotron power and $P_{IC}$ is IC power emitted by an electron).       
However, protons will radiate at a very low rate compared to electrons/positrons ($P_{syn/IC}\propto\frac{1}{m_p^2}$).
So, we can say that the radio lobes contain radiating particles and non-radiating particles in general. Since, 
`whether the radio galaxy lobes are in the minimum energy condition' has always been an issue in the field of radio 
astronomy, astronomers test the minimum energy hypothesis whenever possible. In that case, we need to estimate 
the ratio ($\kappa$) of kinetic energy in non-radiating particles to that of radiating particles. Therefore, particle 
content in radio lobes or at least the knowledge of $\kappa$ is a very important issue.
From recent studies of large radio galaxies ($l=$ a few 100 kpc to Mpc), it has been found that the lobes are close
to the minimum energy condition (Croston et al. 2004, 2005, Konar et al. 2009) with the assumption that the non-radiating 
particles are not energetically dominant. Since, the outer lobes of DDRGs have been created in the same way
as the lobes of radio galaxies, those outer lobes are likely to have similar physical properties. But the density 
in relativistic plasma in the old cocoons into which the inner lobes are 
expanding, if equipartition holds, is orders of magnitude below that needed to ram-pressure confine the inner lobes
even if we take protons (as non-radiating particles) into consideration. In other words, in such a tenuous medium 
the inner jets should have moved ballistically, had there been only relativistic matter fed by the jets into outer 
lobes in previous cycle. But we clearly see well-defined inner lobes embedded in the outer lobes of the DDRGs.
This compels us to think that besides electron-positron/proton fed by the jets, thermal matter from the ambient medium
must have been ingested into the outer lobes. Therefore, we can say that we would not expect to observe the inner lobes 
unless some additional, higher density material is present in the outer cocoons. 
Kaiser, Schoenmakers \& R\"{o}ttgering (2000) argued that entrainment of material from surrounding hot gaseous environment 
is not a viable source of material to confine the inner lobes, as it is a very slow process. Instead, they proposed 
a two fluid scenario in which dense clumps of warm ($10^4$ K) gas embedded in the intragroup/intracluster medium are 
ingested inside the expanding lobes and mixed in with the cocoon material after the clumps get disrupted. It is possible
that those warm clumps get heated up by the interaction with the relativistic cocoon material.
For the cases where thermal material might provide the missing pressure (e.g., in the case of FRI lobes), it has 
to be heated in order to make its density lower so that cavities are observed in the positions of the lobes 
(e.g. Croston et al 2003, Hardcastle, Sakelliou \& Worrall 2005). So, there must be some heating mechanism in order 
to provide the missing pressure in the FRI lobes and this must be provided by the microphysics, which seem unlikely to be
different for FRIs and FRIIs. We do not know what temperature it is heated to, but it must be considerably 
larger than the ambient temperature. Once the entrained material has been heated, it's possible that some processes, 
like shocks, will drive some protons into a non-thermal tail.
The work of Safouris et al. (2008) has shown that there is strong evidence for the ambient thermal matter being ingested 
into the outer lobes of DDRGs. Therefore, the inner jets are propagating through a medium which is in general a 
mixture of relativistic gas and ingested thermal gas from the ambient medium. However, Brocksopp et al. (2007, 2011)
put forward an alternative model according to which the inner lobes arise from the emission of relativistic electrons 
within the outer lobes, which are compressed and re-accelerated by the bow shock in front of the restarted jets and 
within the outer lobes. 

As the discussion in the previous paragraph shows, the issues that are crucial for an understanding of the radio 
galaxy dynamics are 1) the proton to electron energy density ratio $\kappa$, 2) the magnetic field strength, 
3) internal and external pressures of lobes, 4) the ingested thermal matter density inside the lobes and the 
kinetic energy density in the ingested thermal matter, and 5) the actual picture of the formation of inner loles. 
To obtain these pieces of information, it is necessary to study radio galaxies as well as DDRGs both in X-ray 
and radio wavelengths.

\section{How we can study the dynamics of radio galaxies by radio and X-ray observations}
\label{sec:radioxray}
As described above, leptons (e$^-$/e$^+$) in the lobes radiate via the synchrotron process at radio wavelengths. 
We know that Cosmic Microwave Background (CMB) photons are ubiquitous in the Universe. So the same 
charged particles, which radiate via synchrotron, also up-scatter CMB photons from millimeter wavelengths
to X-ray wavelengths. We will hereafter refer to this IC scattering of CMB photons as the IC-CMB process. 
In addition, there is an ambient thermal medium in which a radio galaxy is always embedded. Such a medium 
radiates via the thermal bremsstrahlung in the X-ray. If the radio lobes are strong enough and the ambient thermal 
medium is relatively weak, we may be able to measure the X-ray from the lobes. The detection of lobe related X-ray 
depends on how well the background subtraction can be done. Once we detect such lobe related X-ray, we can try to fit
the data with IC-CMB and thermal bremsstrahlung models. The lobe-related X-rays detected so far in all radio 
galaxies are consistent with IC-CMB process, though the thermal origin of lobe related X-ray cannot be ruled out
in some cases. In some cases, e.g, Pictor A (Hardcastle \& Croston 2005) it is clearly ruled out. As an example, 
we describe here radio and X-ray observational results of a giant radio galaxy, 3C457, in brief in 
Section~\ref{sec:GRG3C457} as a case study. 

Since we are interested in the dynamics of the radio lobes, age is an important 
parameter of the source. We use spectral ageing analysis for estimating the ages. Caveats related
to the evolution of the local magnetic field in the lobes need to be borne in mind (e.g. Rudnick et al. 1994; 
Jones et al. 1999; Blundell \& Rawlings 2000). Furthermore, while Kaiser (2000) have suggested that spectral 
and dynamical ages are comparable if bulk back-flow and both radiative and adiabatic losses are taken into
account in a self-consistent manner, Blundell \& Rawlings (2000) suggest that this may be so only in the 
young sources with ages much less than 10 Myr. In the study of the FR II-type giant radio galaxy, J1343+3758, 
Jamrozy et al. (2005) find the dynamical age to be approximately four times the maximum synchrotron age of
the emitting particles. In spite of the caveats of the spectral ageing analysis, we get a reasonably good 
idea about the dynamics of the sources by using spectral ages. Spectral ageing analysis not only infers the ages, 
but also tell us whether the particles are re-accelerated in certain portions of the lobes (after being accelerated 
at the hotspot during injection). This hints us some physical processes (possibly lobe-environment interactions) 
which are not clearly known.   

\subsection{The giant radio galaxy 3C457: a case study}
\label{sec:GRG3C457}
\begin{figure}[h!]
\hbox{
\epsfig{file=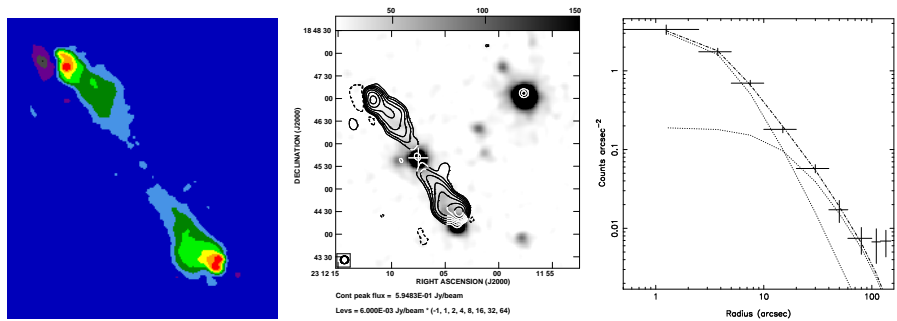,height=4.0cm,angle=0}
\epsfig{file=3C457_G_PSC4+3C457XMM_COR.PS,height=4.5cm,angle=0}
\epsfig{file=proplot_out.ps,height=4.0cm,angle=0}
}
\caption{Left panel: A false colour GMRT map of the giant radio galaxy 3C457 at 610 MHz at full resolution 
($\sim$6$^{\prime\prime}$). 
Middle panel: X-ray image, convolved with 10$^{\prime\prime}$ Gaussian, of the source is in grey scale 
which is overlaid in contour with a 610-MHz GMRT image of 10$^{\prime\prime}$ resolution. A `$+$' sign indicates
the position of the optical host galaxy.
Right panel: The surface brightness profile of the environment of 3C\,457.
The dot-dashed line is the total model (a point source plus a
$\beta$-model) fitted to the data from all three cameras and dotted
lines are two components of the total model. For detail, see Konar et al. (2009), from which
the middle and the right panels have been reproduced. }
\label{3C457_map1}
\end{figure}

This particular source (as shown in Fig.~\ref{3C457_map1}) is a giant radio galaxy of size $\sim$ 1 Mpc at a 
redshift of 0.428. Our detailed work on this source has been published by Konar et al. (2009). Fig.~\ref{3C457_map1} 
shows, both in colour and contour, the GMRT images depicting diffuse relativistic plasma filling the huge lobes of 
this source. At the ends of the lobes we can see the clear double hotspots, which might have been created due 
to jittering of the jet direction. These hotspots are the characteristic features of FRII (Fanaroff \& Riley, 1974) 
radio galaxies and the sites of particle acceleration. 
IC-CMB modelling of the lobes yields that the magnetic fields ($B$) are 0.68$^{+0.06}_{-0.09}$ and
0.40$^{+0.04}_{-0.02}$ nT for the NE and SW lobes respectively (the internal pressure of the lobes are
$0.224^{+0.034}_{-0.019}\times10^{-12}$  and $0.378^{+0.043}_{-0.042}\times10^{-12}$ Pa for the NE and
SW lobe respectively). The corresponding minimum energy fields ($B_{min}$) are 0.86 and 0.82 nT, which are within a 
factor of two of what we measure. This is in good agreement with the results of Croston et al. (2004, 2005). 
Minimum energy fields are calculated with the assumptions that the filling factor of lobes is unity, the 
energetically dominant particles are the radiating particles only 
(the contribution of protons has been neglected, i.e., $\kappa=0$), and the electron energy spectra extend 
from $\gamma=$10 to 10$^{5}$. Croston et al. 2005 argue that the matching of the values between 
$B$ (constrained from IC-CMB modelling) and $B_{min}$ (estimated with such a choice of parameters) has to 
be a coincidence for all radio lobes, unless these parameters have values close to what have been assumed.  
However, such a coincidence in all the lobes studied by Croston et al. (2004, 2005) and Konar et al. (2009)
is unlikely. So in that sense, the IC-CMB modelling of radio lobes (including the lobes of 3C457) says that 
the non-radiating particles are not energetically dominant, i.e., $\kappa \sim 0$.  Our spectral ageing 
analysis of the same source has yielded an age of $\sim$30 Myr.  

The middle panel of Fig.~\ref{3C457_map1} shows that the lobes as well as the environment are detected in X-ray.
The right panel of Fig.~\ref{3C457_map1} shows the X-ray surface brightness profile of the environment. We fitted the 
spectrum of the environment with the mekal model (thermal bremsstrahlung model) at a redshift of 0.428 
(see Konar et al. 2009 for details of the fit). This yields a reasonable fit with the gas temperature equal 
to 2.62$^{+1.15}_{-0.69}$ keV. Our estimated luminosity of the hot gaseous environment is 
$\sim$1.51$^{+1.33}_{-0.97}\times$10$^{44}$ erg s$^{-1}$. This is a poor cluster scale environment as we 
find from the luminosity$-$temperature correlation (Osmond \& Ponman 2004).

\begin{figure}[h!]
\hbox{
\epsfig{file=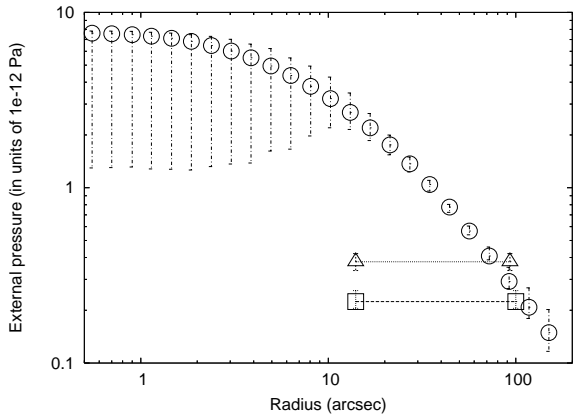,height=8.0cm,angle=-90}
}
\caption{
Pressure vs. deprojected radius plot of the environment of
         3C\,457 at the best-fitting temperature of 2.62 keV. The
         deprojected radius (on the x-axis) is measured from the
         position of the core of 3C\,457. Open circles: pressure of the
         environment vs. radial distance from the core. Open squares:
         pressure of the NE lobe vs. radial distance from the core.
         Open triangles: pressure of the SW lobe vs. radial distance
         from the core. The open triangle and the open square on the 
         right hand side indicates the positions of the hotspots of
         the two respective lobes. The lobes are assumed to be in the plane of
         the sky in this plot. The error in pressure shown in this
         plot is only due to the error in measurement of the emission
         measure only. Errors on temperature are not shown for clarity.
         This figure has been reproduced from the work of Konar et al. (2009).
}
\label{pressure.profile}
\end{figure}

The radial surface brightness in units of counts arcsec$^{-2}$ was extracted from concentric annuli with 
point sources, chip gaps and lobes masked out (see Fig.~\ref{3C457_map1}). 
We fitted a model consisting of a point source situated at 
the position of the core of 3C\,457 and a single $\beta$ model. The Bayesian estimates of $\beta$ and core 
radius ($r_c$) are $\beta = 0.51^{+0.02}_{-0.17}$ and $r_c = 11.2_{-6.4}^{+22.7}$ arcsec. Errors are the 1-$\sigma$ 
credible interval. The detailed analysis of the spatially extended emission around the core of 3C\,457 and 
the $\beta$-model fitting procedure were carried out in the same way as described in Croston et al. (2008).
Knowing the temperature and the surface brightness profile of the ambient medium, we have estimated the pressure 
profile. Fig.~\ref{pressure.profile} shows the pressure profile of the hot gaseous ambient medium of 3C\,457. 
Now it is evident from  Fig.~\ref{pressure.profile} that the lobe heads are close to pressure balance with
the ambient medium. The middle and the tail-part of the lobes are little underpressured, which is in contradiction
with what has been assumed in self-similar models of radio galaxy (Falle 1991; Kaiser \& Alexander 1997; 
Kaiser, Dennett-Thorpe \& Alexander 1997). We believe that the lateral expansion is subsonic, as otherwise; 
we would not have been able to model the lobe related X-ray as due to pure IC-CMB process. So, the results 
are consistent with our paradigm of radio galaxy dynamics with radiating particles as energetically dominant.\\

Another quantity which is important for understanding the dynamics of radio galaxies is the lobe head velocity.
Since the lobe heads of radio galaxies propagate through the dense thermal ambient medium, lobe head velocity 
is non-relativistic. So the ram pressure balance equation at the lobe head gives us the following
equation.
\begin{equation} 
n_p\mu m_p v_h^2 = \frac{L_{jet}}{v_{jet}A_h},
\label{ram.pressure}
\end{equation}
where $n_p$ is the particle density of the environment, $\mu$ is the mean molecular weight of the particles 
in the environment and can be assumed to be 1.4, $m_p$ is the proton mass, $v_h$ is the jet head velocity, 
$L_{jet}$ is the jet power and can be expressed as $\frac{U_{tot}}{t}$ with t = spectral age, $v_{jet}$ is the 
jet bulk speed which is usually assumed to be close to the speed of light and $A_h$ is the lobe head area over 
which the jet is impinging the environment. All the parameters except $A_h$ in Equation~\ref{ram.pressure} have 
unique values for the source 3C457. However, for $A_h$, we have estimated $\sim$2 kpc and $\sim$25 kpc
as lower and upper limits of the diameters of the hotspots for 3C457. Our estimation yields that $v_h$ is 
equal to $\sim$0.04$c$ (for both lobes) and 0.003$c$ (for both lobes) for lower and higher limits of $A_h$ 
respectively. So, the lobe heads can be supersonic, though the lateral expansion is subsonic, which is consistent 
with our IC-CMB modelling of the lobe related X-rays. Hence the entire analysis of the dynamics of FRII radio galaxies
is self consistent.

All our calculations presented here are with the assumption that the lobes are on the plane of the sky. 
Since 3C457 is a radio galaxy, it can be aligned at an angle between 45$^{\circ}$ and 90$^{\circ}$ with
the line of sight. If we repeat our calculation with the assumption that our sources are aligned at an 
angle of 45$^{\circ}$ with the line of sight, then the values of magnetic fields and internal pressure 
of lobes don't change much. Hence the qualitative results related to the pressure balance of lobes remain 
more or less the same.

\section{Double-double radio galaxies}
\label{sec:ddrg}
In the case of DDRGs, the inner lobes are often confined within the cocoon material of the outer lobes.
Illustrative examples of such sources are J0041$+$3224, J1548$-$3216, J0116$-$4722 
(Saikia et al. 2006; Safouris et al. 2008; Saripalli et al. 2003; Saripalli et al. 2002).
\begin{figure}[h!]
\hbox{
\epsfig{file=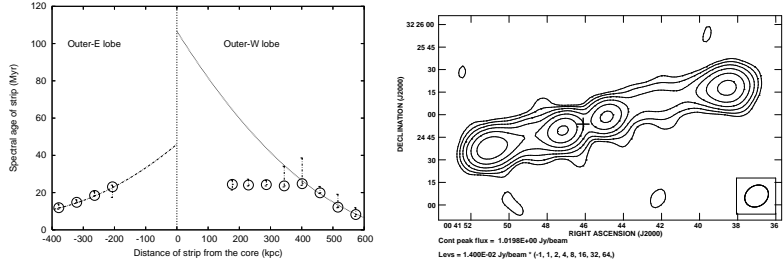,height=5.0cm,angle=-90}
\epsfig{file=J0041+3224_T.PS,height=5.3cm,angle=-90}
}
\caption{Left panel: A plot of spectral age vs. strip distance from the core.
Right panel: 240 MHz GMRT image of DDRG J0041$+$3224. A `$+$' sign indicates the position of the 
optical host galaxy. 
}
\label{specage_j0041}
\end{figure}
\begin{figure}[h!]
\hbox{
\epsfig{file=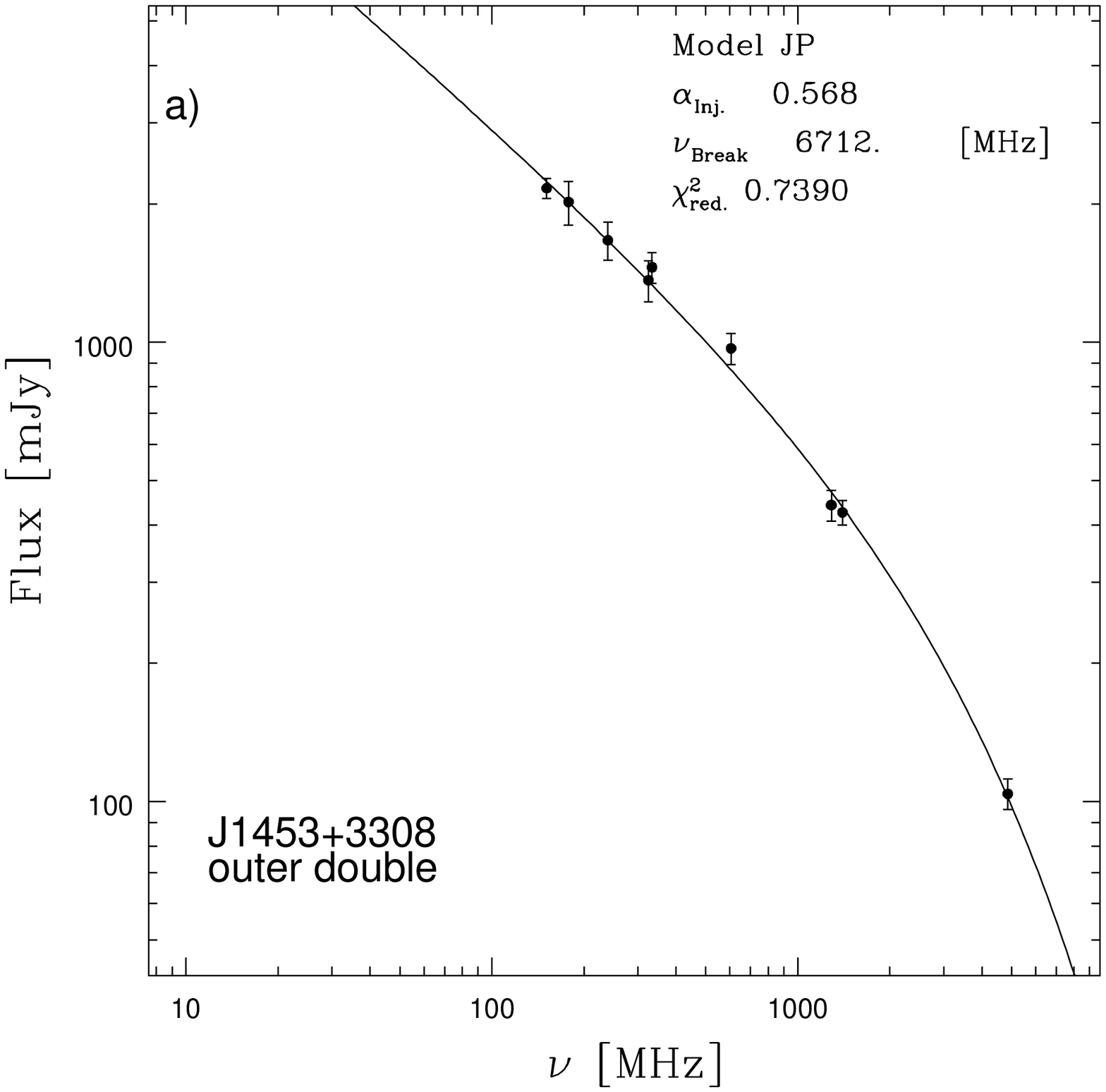,width=4.5cm,angle=0}
\epsfig{file=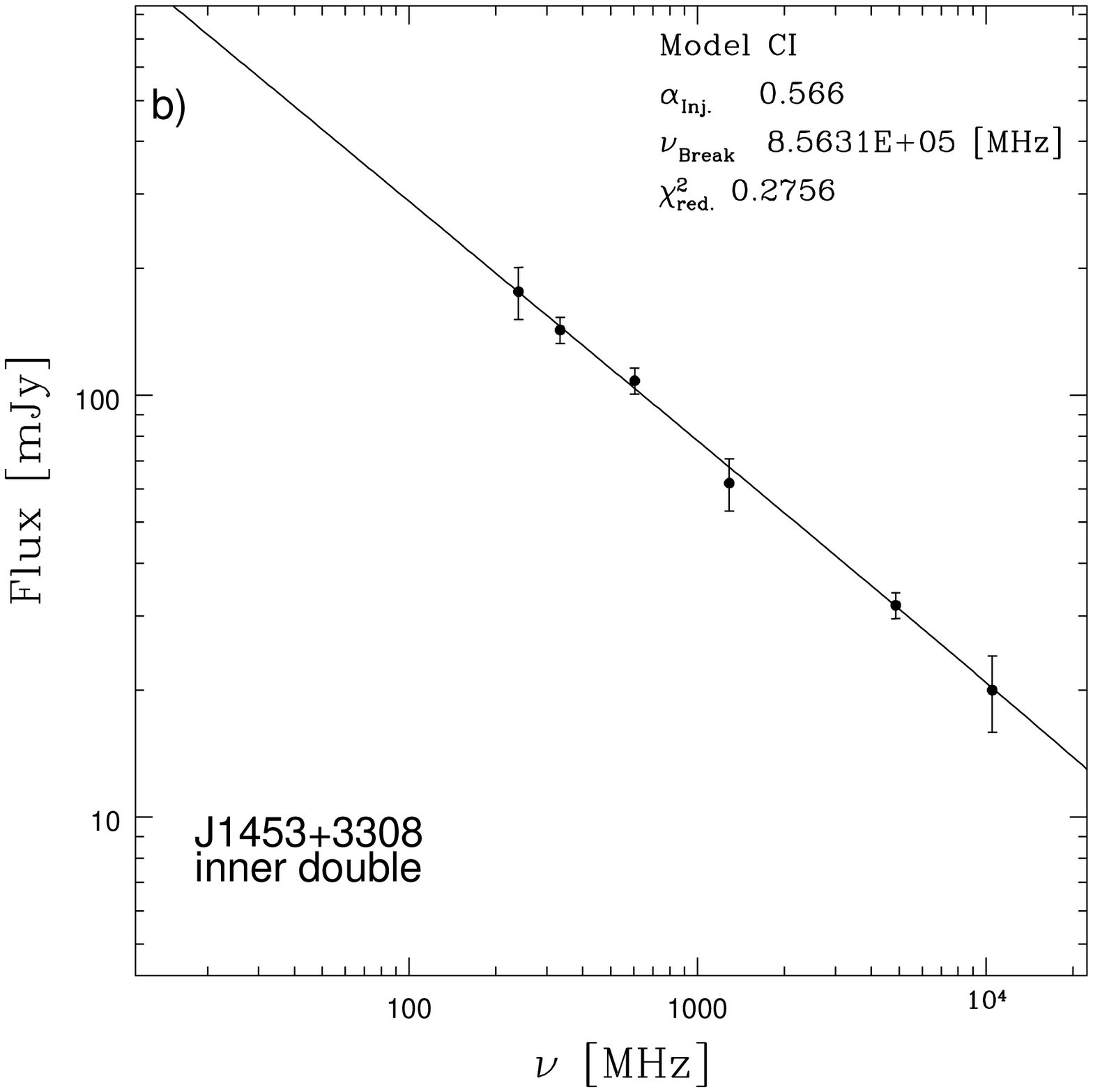,width=4.5cm,angle=0}
\epsfig{file=J1453+3309_610.PS,width=2.0cm,angle=0}
}
\caption{Left panel: The total spectrum of the outer lobes of J1453$+$3308. Middle panel: The total spectrum of 
the inner double (excluding the core) of the same source. Right panel: 610 MHz GMRT image of the same source.
These are reproduced from the work of Konar et al. (2006).}
\label{spect.j1453}
\end{figure}

We have made  multifrequency radio observations of a small sample of DDRGs to study their dynamics.
Preliminary results for the source J0041$+$3224 (see Fig.~\ref{specage_j0041}) are presented here. From our 
spectral ageing analysis of J0041$+$3224, we found, in the spectral age vs. distance plot as presented 
in Fig.~\ref{specage_j0041}, that the age is increasing monotonically with the distance of the strip 
from the hotspot for eastern lobe. However, for the western lobe, the spectral age increases with the 
distance from the hotspot up to 200 kpc from the hotspot. 
After that the spectral age does not vary with the distance from the hotspot. This we interpret as due to particle 
re-acceleration in that portion of the western lobe. So, this result hints at some unknown physical processes 
causing particle re-acceleration in the lobes. We have found that the average injection spectral index, $\alpha_{inj}$ 
($S_{\nu} \propto \nu^{-\alpha_{inj}}$) is similar ($\sim$0.76) for both the outer as well 
as inner double for J0041$+$3224. 
We also found similar $\alpha_{inj}$ values, in both episodes of jet activity, of $\sim$0.57 and 0.80 for J1453$+$3308 
(see Fig.~\ref{spect.j1453}, Konar et al. 2006) and 4C29.30 (Jamrozy et al. 2007) respectively. Readers can notice 
the similarity of the slopes of two power-law spectra (spectrum of inner double and low frequency part of the outer 
double spectrum) of J1453$+$3308 (Fig.~\ref{spect.j1453}). The outer double spectrum (Left panel, Fig.~\ref{spect.j1453}) 
of J1453$+$3308, has been fitted with Jaffe-Perola model (Jaffe \& Perola 1973). All these results will be discussed 
in more detail by Konar et al. (2012a), Konar et al. (2012b in prep) and Konar \& Hardcastle (2012, in prep).

\section{Summary of the results from the study of the dynamics of radio galaxy and double-double radio galaxies}
The results that we have obtained from our study of the dynamics of 3C457 are as follows.
\begin{enumerate}
\item The magnetic fields of radio lobes are close to the minimum energy values.
\item Non-radiating particles in lobes do not seem to be energetically dominant.
\item The lobes appear to be approximately at pressure balance towards the edge and underpressured towards the core.
\item In our paradigm, the lobe heads can be supersonic, though the lateral expansion is subsonic. 
\item There can be re-acceleration of particles in lobes (other than hotspot regions), which hints at some physical 
      process (possibly lobe-environment interaction) inside the lobes.
\item The injection spectral index seem to be similar in two episodes of jet activity, which hints 
      that intrinsic source properties should determine the values of $\alpha_{inj}$.
\end{enumerate}

\section*{Acknowledgments}
We thank the staff of the GMRT, VLA and XMM-Newton for the observations. This research has made use of the NASA/IPAC 
extragalactic database (NED) which is operated by the Jet Propulsion Laboratory, Caltech, under contract with the 
National Aeronautics and Space Administration. We thank numerous contributors to the GNU/Linux group.
MJ is partially supported by MNiSW through the research project 3812/B/H03/2009/36 during the years 2009-2012.


\label{lastpage}
\end{document}